\documentclass[prb,twocolumn,showpacs,amsmath,amssymb]{revtex4}

\usepackage{graphicx}
\usepackage{dcolumn}
\usepackage{bm}
\usepackage{amsmath}
\usepackage{epstopdf}
\usepackage{tipa}
\usepackage{upgreek}
\usepackage{subfigure}
\usepackage{mhchem}

\newcommand{\sgn}{\text{ sgn$\mkern2mu$}}
\def\im{\,\mathrm{Im}}
\def\det{\mathrm{det} \,}

\usepackage[mathlines]{lineno}

\begin{document}

\preprint{APS/123-QED}

\title{Interaction between domain walls in chiral $p$-wave superconductors}
\author{M. A. Przedborski}
\email{mp06lj@brocku.ca}
\author{K. V. Samokhin}
\email{kirill.samokhin@brocku.ca}
\affiliation{Physics Department, Brock University, St. Catharines, Ontario, Canada L2S 3A1 }

\date{\today}

\begin{abstract}
We calculate microscopically the interaction energy of domain walls separating degenerate ground states in a chiral $p$-wave superconductor. The interaction is mediated by the quasiparticles experiencing Andreev scattering at the domain walls. 
As a by-product, we derive a useful general expression for the free energy of an arbitrary nonuniform texture of the order parameter in terms of the quasiparticle scattering matrix.  
\end{abstract}

\pacs{74.20.Rp, 74.20.-z, 67.30.hp}

\maketitle

\section{\label{sec:intro} Introduction}

Recent years have seen an increase in interest in topological superconductors and superfluids from both experimental and theoretical facets,~\cite{top-SC} with one of the most studied examples being the chiral $p$-wave triplet state. 
The defining property of these systems is that, while the fermionic excitations in the bulk are fully gapped, there are gapless quasiparticles, which are protected by topology and are localized near inhomogeneities of the order parameter, 
such as sample boundaries, domain walls (DWs), and Abrikosov vortices. The chiral $p$-wave state in particular has received notable consideration because of the gapless quasiparticle excitations (Majorana fermions) and non-Abelian winding statistics associated with half-quantum vortices, which are potentially useful as a route to quantum computing.~\cite{KI01}

The chiral $p$-wave triplet pairing, with order parameter proportional to $k_{x}\pm i k_{y}$, is experimentally realized in the superconducting state of \cf{Sr2RuO4} (Refs.~\onlinecite{MM03} and \onlinecite{KB09}), 
as well as the A-phase of superfluid $^{3}$He (Ref.~\onlinecite{VW02}). The ground state is two-fold degenerate in the absence of an external magnetic field, and this gives rise to the possibility of superconducting or superfluid states with opposite chirality, separated by DWs, to form in different parts of the system.~\cite{VG85,SU91} There is in fact experimental evidence of the existence of DWs in \cf{Sr2RuO4} from Josephson measurements,~\cite{KK08,KS06} and also in thin films of $^{3}$He-A from torsional oscillator measurements.~\cite{WW04} In general, pairing states in other unconventional superconductors can exhibit discrete degeneracies of the ground state,~\cite{VG85,SU91} which also leads to the possibility of DW formation in these systems. 

The formation of a DW costs gradient energy to the system due to the spatial variation of the order parameter. Unlike ferromagnets, which break up into domains in order to minimize the net magnetic moment, in a neutral superfluid there is no analogous energetic rationale behind the formation of DWs. One possible explanation is that the DWs are spontaneously formed due to sample inhomogeneities during cooling across the phase transition into the superfluid state. Alternatively, the creation of low-energy quasiparticles bound to the DW may compensate for the increase in gradient energy, which is particularly effective in one-dimensional systems.~\cite{KY02}    

The purpose of this paper is to develop a general microscopic formalism for calculating the interaction between superconducting DWs at arbitrary temperature. 
The structure of a single DW was investigated in Refs.~\onlinecite{VG85,SU91}, and \onlinecite{DWStructure}. In general, the structure of superconducting DW textures can be studied using the Ginzburg-Landau (GL) formalism. It turns out that there are no stable two-DW solutions, and consequently, there must be some form of interaction between the DWs. Therefore, either an attraction between two DWs must cause an effective collapse of the DWs to a single domain; or a repulsion between them pushes one of the DWs to infinity, leading to the effective formation of just two domains. It is this interaction which has stimulated our current work.

The paper is organized as follows: In Sec.~\ref{sec:2}, we introduce the two-DW configuration, for which we will ultimately determine the interaction energy. In Sec~\ref{sec:3}, we compute the quasiparticle spectrum of this texture in the semiclassical (Andreev) approximation. Then, using the transfer matrix method, we relate the interaction energy to the scattering matrix of the Bogoliubov quasiparticles in Sec~\ref{sec:4}. Finally, we analytically calculate the interaction energy between DWs in the limit of large DW separation. Throughout this paper, we use the units in which $\hbar=k_{B}=1$.

\section{\label{sec:2} The Model}

We consider a two-dimensional chiral $p$-wave neutral superfluid. Any external fields and disorder are neglected and an isotropic Fermi surface is assumed. The order parameter of a triplet fermionic superfluid or superconductor is a $2\times 2$ spin matrix which has the form 
$\hat{\Delta}({\bm k},{\bm r})= i\hat{\sigma}_{2}\hat{\bm{\sigma}}\bm{d}({\bm k},{\bm r})$, where $\hat{\bm{\sigma}}$ are the Pauli matrices and ${\bm d}$ is the spin vector. For unitary states the latter defines the normal to the plane in which fermions paired at $({\bm k},-{\bm k})$ are equal spin paired.\cite{MM03}
In our case, ${\bm d}$ has only $\hat z$-component, and its momentum dependence is given by\cite{Book}
\begin{equation}
	\label{eq:dvec}
	{\bm d}=\frac{\eta_{1}k_{x}+\eta_{2}k_{y}}{k_{F}}\hat z, 
\end{equation}
where $\eta_{1}$ and $\eta_{2}$ are components of a complex order parameter vector ${\bm \eta}$ and $k_F$ is the Fermi wave vector. 

We focus on planar superconducting textures describing one or more DWs perpendicular to the $x$ axis, therefore only $x$-dependence is retained in $\bm{\eta}$. The DWs separate states of opposite chirality, hence
the order parameter alternates between $k_x+ik_y$ and $k_x-ik_y$ states. The spatial dependence of $\bm{\eta}$ can be studied using, e.g. the GL formalism, see Appendix. There is no exact analytical solution for the DW structure, even in the case of a single DW, 
and a variety of approximations have been proposed in the literature (Refs.~\onlinecite{VG85,SU91}, and \onlinecite{DWStructure}). 
Most important qualitative features of the DW textures can be illustrated using the constant-amplitude model introduced by Volovik and Gor'kov in Ref. \onlinecite{VG85}. 
In this model the order parameter has the form $\bm{\eta}(x) = \Delta_{0}(1,e^{-i\gamma(x)})e^{i\phi(x)}$, where $\phi$ is the common phase, and $\gamma$ is the relative phase, of the order parameter components. 

The phases $\phi$ and $\gamma$ are not independent. Conservation of current requires that the transverse current is constant, and since it is fixed by external sources, one may set it to zero. This results in a linear relationship between the gradients of $\phi$ and $\gamma$. Thus the DW texture can be described in terms of a single variable -- a spatially-dependent relative phase $\gamma(x)$. Variational minimization of the GL free energy functional with respect to $\gamma$ leads to a sine-Gordon equation, whose simplest nontrivial solution corresponding to a single DW has a kink-like form, connecting the asymptotics $\gamma(\pm\infty)=\pm\pi/2$ and varying within a region of thickness $\xi_d$ (which has the meaning of the DW thickness). In general, different models give different expressions for $\bm{\eta}(x)$, but the the condition $\gamma(\pm\infty)=\pm\pi/2$ always holds without reference to a specific profile of the order parameter near the wall. Furthermore, the common phase
difference between the two domains is fixed by the condition of vanishing supercurrent across the DW, see Appendix for details. Thus, one can write the order parameter asymptotics far from the single DW as follows:	
\begin{align}
	\label{eq:OP+}
	\bm{\eta}(x)&=\Delta_{0}(1, i), \\
\intertext{at $x \rightarrow -\infty$, and} 
	\label{eq:OP-}
	\bm{\eta}(x)&=\Delta_{0}e^{i\chi}(1,- i), 	 
\end{align}
at $x \rightarrow +\infty$. Here $\chi$ is a parameter depending on the microscopic details of the system, satisfying the condition $0\leq \chi \leq \pi$.  One can make analytical progress by considering the sharp DW model, 
in which case $\xi_{d} \rightarrow 0$ and the order parameter changes abruptly at $x=0$ between its asymptotic values. 

\begin{figure}
	\begin{center}
	\includegraphics[width=6.0cm]{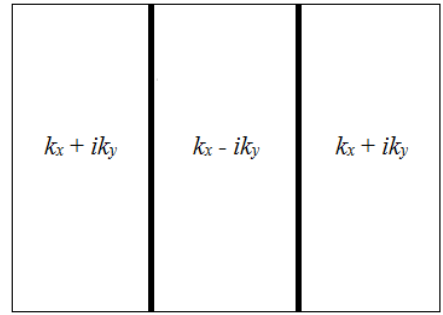}
	\end{center}
	\caption {Alternating chirality states in the two-DW model.}	
	\label{fig:op}
\end{figure}

We now consider two DWs at a fixed separation $L$, with the first DW positioned at $x=0$, and the second at $x=L$. Using a similar setup as in the single DW case, 
the chirality alternates between the three domains, and we analogously impose the constraint of vanishing supercurrent along the $x$ axis, which leads to a non-zero common phase difference between the domains. 
The outer left region ($x<0$), and the region on the far right ($x>L$), correspond to the  $k_{x}+ik_{y}$  state, while the middle domain ($0<x<L$) corresponds to the $k_{x}-ik_{y}$ state, as shown in Fig.~\ref{fig:op}. 

\begin{figure}
	\begin{center}
    	\includegraphics[width=6.0cm]{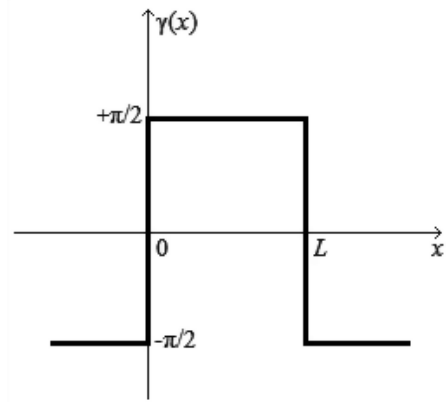}
	\end{center}
	\caption{The relative phase between the order parameter components for two sharp DWs with fixed separation $L$.}	
	\label{fig:gamma}
\end{figure}

As in the single DW case, we focus on the sharp DW model to obtain an analytical solution for the interaction energy of the two DWs. Then the order parameter for both of the outer domains is given by the expression in Eq.~(\ref{eq:OP+}), 
and a non-zero global phase factor appears in the order parameter of the middle domain, 
which is given by the expression in Eq.~(\ref{eq:OP-}). In accordance with the sharp DW model, $\gamma(x)$ changes abruptly between its asymptotic values in the three domains, as illustrated in Fig.~\ref{fig:gamma}.

\section{\label{sec:3} Quasiparticle Spectrum}

Since we consider a neutral superfluid, interaction between the DWs can only be due to their effect on the Bogoliubov fermionic quasiparticles.
The quasiparticle spectrum for a nonuniform superconductor is determined by the Bogoliubov-de Gennes (BdG) equations, with the $4\times 4$ BdG Hamiltonian given by
\begin{equation}
\label{eq:HBdG}
	{\cal H}=\left(\begin{array}{cc}
		\hat\xi & \hat\Delta\\
		\hat\Delta^\dagger & -\hat\xi
	\end{array}\right),
\end{equation}
where $\hat\xi=\xi(\bm{k})\hat{\sigma}_{0}$. Here $\hat{\sigma}_{0}$ is the $2\times 2$ identity matrix, and we assume that $\xi(\bm{k})=(\bm{k}^{2}-k_{F}^{2})/2m^{*}$, where $m^{*}$ is the effective mass. 
For the chiral $p$-wave states, we have $\hat \Delta= d_{z}\hat{\sigma}_{1}$, where $d_{z}\propto k_{x} \pm ik_{y}$. This form for the gap function allows the BdG Hamiltonian in Eq.~(\ref{eq:HBdG}), which operates on a four-component 
quasiparticle wavefunction, to be written as a direct sum of two identical $2\times 2$ matrices, denoted by $H_{BdG}$. The four-component wavefunction is thus decoupled into two two-component wavefunctions which satisfy $ H_{BdG} \Psi_{\sigma} = E \Psi_{\sigma}$,
where $\sigma=\pm$, and $H_{BdG}$ is given by
\begin{equation}
	\label{eq:BdGH}	
	H_{BdG}=\left(\begin{array}{cc}
		\xi & d_{z}\\ \noalign{\vskip 1mm} 
		d^\dagger_z & -\xi
	\end{array}\right).
\end{equation}
We should point out that $\sigma$ does not denote the spin projection of the two-component wavefunction; in fact, the components of both spinors have mixed spin projections. From this point on, we may drop the label $\sigma$, whose only effect is to double the degrees of freedom.

For a DW parallel to the $y$-axis, where the order parameter depends only upon $x$, the $y$-dependence of the quasiparticle wavefunction is equivalent to that of a free particle (i.e. a plane wave). 
It can be written as $e^{ik_{y}y}\Psi(x)$, where $\Psi(x)$ satisfies the two-component BdG equations for a given $k_y$:
\begin{equation}
	\label{eq:BdG}
	\left( \begin{matrix}
		\dfrac{\hat{k}_{x}^{2}-k_{0}^{2}}{2m^{*}} & \Delta(x) \\
		\Delta^{\dagger}(x) & -\dfrac{\hat{k}_{x}^{2}-k_{0}^{2}}{2m^{*}}
	\end{matrix} \right) \Psi = E \Psi,
\end{equation}
with $\hat{k}_{x} = -i\nabla_{x}$, $k_{0}=\sqrt{k_{F}^{2}-k_{y}^{2}}$, and $\Delta(x)=d_{z}(x)=\eta_{1}(x)(\hat{k}_{x}/k_{F})+\eta_{2}(x)(k_{y}/k_{F})$, see Eq.~(\ref{eq:dvec}).

The DW order parameter $\Delta(x)$ varies slowly on the scale of $1/k_{F}$. Consequently, we can apply the semiclassical (Andreev) approximation~\cite{And64} and seek solutions of the form $\Psi(x)=e^{ik_{x}x}\psi(x)$, where $\psi=(u, v)^{T}$ 
is a slowly varying ``envelope" function with electron-like ($u$) and hole-like ($v$) components. Due to the circular symmetry of the Fermi surface in the $xy$ plane, we have $k_{x}=\pm k_{0}$ for a given $k_{y}$. 
Substituting $\Psi(x)$ of this form into Eq.~(\ref{eq:BdG}) and neglecting terms containing higher-order gradients of $\psi$ we see that the envelope function satisfies the Andreev equations:
\begin{equation}
\label{eq:And}
	\left(\begin{matrix}
		-iv_{F,x}\nabla_{x} & \Delta_{\bm{k}_{F}}(x)\\ \noalign{\vskip 2mm}
		\Delta^{*}_{\bm{k}_{F}}(x) & iv_{F,x}\nabla_{x}
	\end{matrix}\right)\psi=E\psi.
\end{equation} 
The direction of semiclassical propagation of quasiparticles is defined by the Fermi wavevector $\bm{k}_{F}\equiv (k_x,k_y)=k_{F}(\cos\theta,\sin\theta)$, the Fermi velocity is given by $v_F=k_F/m^*$, and $v_{F,x}=v_{F}\cos\theta$.  
The DW order parameter at given $\bm{k}_{F}$  has the form
\begin{equation}
 	\Delta_{\bm{k}_F}(x)=\eta_{1}(x)\cos\theta+\eta_{2}(x)\sin\theta=|\Delta_{\bm{k}_{F}}(x)|e^{i\Phi(x)} , 
	\label{eq:Andreev-OP}
\end{equation}
where $|\Delta_{\bm{k}_F}|$ is the magnitude of the semiclassical order parameter and $\Phi$ is its phase. 

The asymptotics of $\Delta_{\bm{k}_F}(x)$ for the chiral $p$-wave state are fixed by Eqs.~(\ref{eq:OP+}) and~(\ref{eq:OP-}); however, different models for the DW structure, see Sec.~\ref{sec:2}, lead to different forms for the order parameter in the 
vicinity of the DW. In the single sharp DW model, the order parameter is uniform within the domains of opposing chirality, changing abruptly at the boundary $x=0$. Thus we have 
$\Delta_{\bm{k}_{F}}(x) = \Delta_{-}\theta(-x) + \Delta_{+}\theta(x)$, where $\Delta_-=\Delta_0e^{i\theta}$, $\Delta_+=\Delta_0e^{i\chi}e^{-i\theta}$, and $\theta(x)$ is the Heaviside step function. For two DWs, we have 
$\Delta_{-}\theta(-x) + \Delta_{+}\theta(x)\theta(L-x) + \Delta_{-}\theta(x-L)$.

There are two types of solutions supported by the Andreev equations~(\ref{eq:And}): discrete bound states (Andreev bound states, or ABS's) for which $|E|\leq\Delta_{0}$, as well as a continuum of scattering states where $|E|>\Delta_{0}$. It will be shown below that
all quantities of interest, including the interaction between the DWs and also the ABS spectrum, can be expressed in terms of the properties of the scattering states, encoded in the scattering matrix $\hat S$.

\subsection{\label{sec:3a} Scattering matrix for two DWs}

The scattering matrix $\hat S$ relates the amplitudes of the incident wavefunctions to the outgoing amplitudes of the waves that are reflected/transmitted by the DW configuration. To facilitate the calculation of $\hat S$, 
we perform a gauge transformation on the wavefunctions $\psi$ to remove the phase in the order parameter $\Delta_{\bm{k}_{F}}$, see Eq.~(\ref{eq:Andreev-OP}), that appears in the off-diagonal elements of the Andreev Hamiltonian. For the two-DW setup described in Sec.~\ref{sec:2}, we can adopt a more convenient notation to describe the order parameter in each of the three domains:
\begin{equation}
	\label{eq:Delta}
    \Delta_{\bm{k}_F}(x)=\left\{
	    \begin{array}{l}
      		\Delta_{0}e^{i\varphi_{1}},\qquad  x<0,\mkern6mu x> L \\
      		\Delta_{0}e^{i\varphi_{2}},\qquad 0<x<L  
	    \end{array}\right . , 	
\end{equation}
which can be further simplified to $\Delta_{\bm{k}_F}(x)=\Delta_{0}e^{i\Phi(x)}$, with $\Phi=\varphi_{2}=\chi-\theta$ in the middle domain, and $\Phi=\varphi_{1}=\theta$ in the outer two domains. 

We denote the gauge-transformed wavefunctions by $\tilde \psi(x)$ and define $\psi=\hat{U}\tilde \psi$, where $\hat{U}$ is given by $\hat U = e^{i\Phi(x)\hat{\sigma}_{3}/2}$.
The gauge-transformed wavefunctions satisfy $\tilde\psi(+\infty)=\tilde\psi(-\infty)$, which follows from the condition $\psi(+\infty)=\psi(-\infty)$. The latter is consistent with the gap equation and the 
order parameter asymptotics $\Delta_{\bm{k}_F}(+\infty)=\Delta_{\bm{k}_F}(-\infty)$.

Continuity of the original wavefunctions at the boundaries $x=0,  L$ implies $\psi(+0)=\psi(-0)$ and $\psi(L+0)=\psi(L-0)$; however, removing the phase from the order parameter causes the gauge-transformed wavefunctions $\tilde\psi$ 
to suffer a phase discontinuity at the domain boundaries. In fact, one can easily verify that $\tilde\psi$ satisfy the following conditions:
\begin{equation}
\label{eq:BC}
    \begin{aligned}
    &\tilde \psi(+0)= e^{-i\delta}\tilde \psi(-0),\\
    &\tilde \psi(L+0)= e^{i\delta}\tilde \psi(L-0),
    \end{aligned}
\end{equation}
where $\delta=(\varphi_2-\varphi_1)/2=\chi/2-\theta$. 

Although we remove the phase $\Phi$ from the off-diagonal terms in the Andreev Hamiltonian, its derivative $\Phi'$ appears in the diagonal elements after the gauge transformation, so that $\tilde\psi$ satisfies the equation
\begin{equation}
\label{eq:And-gt}
	\left[-iv_{F,x}\hat\sigma_3\nabla_{x} + \frac{1}{2}v_{F,x}\Phi'(x)\hat\sigma_0 +\Delta_{0}\hat\sigma_1\right]\tilde\psi=E\tilde\psi,
\end{equation} 
with the same energy eigenvalues as the original wavefunctions. Due to the delta-function singularities of $\Phi'$ at $x=0,L$, we apply the gauge transformation separately in each region (i.e. at all $x\neq 0, L$, where $\Phi'=0$). In this way we obtain
an equation in each domain which has the form of the Andreev equation in a uniform superconductor, and as such can easily be solved. The solution must satisfy the ``twisted'' matching conditions given by Eq. (\ref{eq:BC}). 

We focus on the scattering states and from now on drop the tilde on the gauge-transformed wavefunctions. For the continuum of scattering states, the quasiparticle wavefunctions are linear combinations of plane waves:
\begin{equation}
\label{eq:QPwf}
	\psi(x)=\sum_{\alpha=\pm} A_{\alpha}e^{\alpha i q x}
           \left(\begin{matrix}
		u_{\alpha} 	\\  \noalign{\vskip 1mm} 
		1
	\end{matrix}\right)
\end{equation}
where $u_{\pm}=\Delta_{0}/(E\mp qv_{F,x})$, with $q =\sqrt{E^2-\Delta_{0}^2}/|v_{F,x}| \\ 
\geq 0$. The subscript on the amplitudes in Eq.~(\ref{eq:QPwf}) corresponds to the direction of 
quasiparticle propagation (i.e. left or right). We also introduce a superscript on the amplitudes of the wavefunctions in the outer two domains ($x<0, x>L$) to identify each particular region. 
Let the ``$-$" superscript denote the region $x<0$, and the ``$+$" superscript correspond to the $x>L$ region. Then the amplitudes of the waves incident on the DW configuration are given by $A_{+}^{(-)}$ and $A_{-}^{(+)}$, 
while the outgoing (reflected and transmitted) waves have amplitudes $A_{+}^{(+)}$ and $A_{-}^{(-)}$. We define the scattering matrix $\hat S$ as follows:
\begin{equation}
\label{eq:Scat}
	\left(\begin{matrix}
		A_{+}^{(+)}  \\ \noalign{\vskip 2mm} 
                     A_{-}^{(-)}
	\end{matrix}\right) = \hat S
	\left( \begin{matrix}
		A_{+}^{(-)}  \\ \noalign{\vskip 2mm} 
		A_{-}^{(+)}
	\end{matrix}\right).
\end{equation}

The scattering matrix is calculated by using the matching conditions given by Eq. (\ref{eq:BC}) to eliminate the wave amplitudes in the region $0<x<L$ and relate the amplitudes of incident waves to those of the outgoing waves. 
The final result has the following form:
\begin{equation}
\label{eq:ScatEnt}
    \begin{array}{l}
     S_{11}= S_{22}=\dfrac{1}{P}, \\
     S_{12}=\dfrac{R_-}{P},\quad S_{12}=\dfrac{R_+}{P},
    \end{array}
\end{equation}
where 
\begin{eqnarray*}
 && P = 1-\frac{\Delta_0^2}{E^2-\Delta_{0}^2}(e^{2iqL}-1)\sin^2{\delta},\\
 && R_\pm = \left(\varrho\pm 1\right)\left(i\cos{\delta}\pm\varrho\sin{\delta}\right)\left(e^{\pm 2iqL}-1\right)\sin{\delta},
\end{eqnarray*}
and $\varrho=E/qv_{F,x}$. 

To conclude this subsection we note that, by considering scattering from the left and right separately, one can relate the scattering matrix to the reflection and transmission coefficients of the Bogoliubov quasiparticles 
in the presence of the order parameter texture. If, for example, there is a wave incident on the DW located at $x=0$ from the left, then we can set the incident amplitude $A_{+}^{(-)}=1$, 
and we must have $A_{-}^{(+)}=0$, $A_{+}^{(+)}=t_{L}$ and $A_{-}^{(-)}=r_{L}$, where $t_{L}$ and $r_{L}$ represent the left-incident transmission and reflection coefficients, respectively.  
The right-incident transmission and reflection coefficients $t_{R}$ and $r_{R}$ can be introduced in a similar way, by considering right-incident scattering on the DW at $x=L$. Then one can relate the scattering matrix entries
to the transmission and reflection coefficients as follows: $t_L=S_{11}$, $r_L=S_{21}$,  $t_R=S_{22}$, and $r_R=S_{12}$.
We should point out that since we have applied the gauge transformation to the quasiparticle wavefunction before calculating the scattering matrix, the reflection and transmission coefficients 
obtained by this method are not equivalent to the ones that would be obtained from the direct Andreev calculation (prior to the gauge transformation).

\subsection{Bound state spectrum}
\label{sec:ABS}

The ABS energy for the single sharp DW model has been previously calculated,\cite{Samo} and it has the following form:
\begin{equation}
	\label{eq:E-bound}
	 E_0(\theta)=\Delta_0s(\theta)\cos\left(\theta-\frac{\chi}{2}\right),
\end{equation}
with $s(\theta)=\sgn\left[\sin\left(\theta-\chi/2\right)\cos\theta\right]$. This expression is valid for an arbitrary phase difference across the DW and it should be noted that, in general, the ABS energy is not a continuous function of $\theta$. 
There are certain directions of semiclassical propagation at which discontinuities occur: at $\theta=\pm\pi/2$, corresponding to a ``grazing trajectory" where the quasiparticles move parallel to the DW (in this case the Andreev approximation is actually not applicable); and also at $\theta=\chi/2$ and $\theta=\chi/2+\pi$, in which case the quasiparticles do not ``see" the DW, since $\Delta_+=\Delta_-$. 
The ABS energies for $\chi=0,\pi$ were calculated in Ref.~\onlinecite{ABS1DW}.  

We proceed now with the calculation of the bound state energies for two DWs. The ABS energies are obtained from the poles in the scattering matrix entries, see Eq. (\ref{eq:ScatEnt}), 
after analytical continuation to the real energy axis within the interval $|E|\leq \Delta_{0}$. Defining the dimensionless energy $\epsilon=E/\Delta_{0}$, we rewrite $q=(\Delta_{0}/v_{F})\sqrt{\epsilon^2-1}/|\cos\theta|$. 
For the bound state energies with $|\epsilon|\leq 1$, we have to choose the correct branch of $q$ before proceeding further. To this end, we consider the function $w(z)=\sqrt{z^2-1}$, which has two branch points: one at $z=1$, and the other at $z=-1$. We choose the branch to ensure that $w(z)$ is real and $w(z)\geq0$ if $z$ is real and $|z|>1$, in accordance with the 
definition of $q$. One can select the branch cuts to run parallel to the imaginary axis, from $\pm\Delta_0$ to $\pm\Delta_0\mp i\infty$, and then the correct choice is $w(z)=i\sqrt{1-z^2}$ for $z$ along the real axis within the interval $[-1,1]$. 

\begin{figure}
  	\includegraphics[width=8cm]{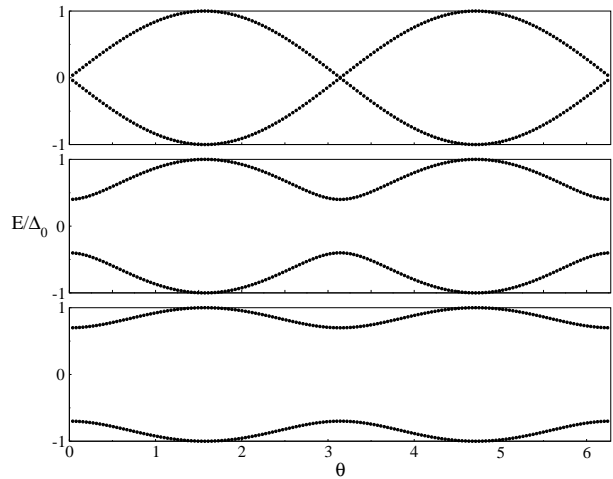}
  	\caption{Bound state energy for the two sharp DW model with $\chi=\pi$, for varying dimensionless DW separation $\tilde L=L/\xi$. From the top: $\tilde L = 5$, $1$, $0.5$.}
  	\label{fig:BSE2DW}    
\end{figure}

To simplify the denominator $P$ in Eq.~(\ref{eq:ScatEnt}), we define $\tilde{\alpha}(\epsilon)=e^{2iqL}=e^{-2\tilde{L}\sqrt{1-\epsilon^2}/|\cos\theta|}$, where $\tilde L=L/\xi$ is the dimensionless distance between the DWs and $\xi=v_{F}/\Delta_{0}$ is the correlation length. Then from the poles in the scattering matrix entries we obtain the following equation for the bound state energies:
\begin{equation}
	\label{eq:BSE2DW}
	\epsilon^2-\cos^2\delta-\tilde\alpha(\epsilon)\sin^2\delta=0.
\end{equation}
To make analytical progress, we can consider this last equation for small values of $\tilde \alpha$, which physically corresponds to large DW separation. Expanding the equation in powers of $\tilde \alpha$ up to linear order, we obtain the result 
$\epsilon=\pm|\cos\delta|[1+(\tilde\alpha\tan^2\delta)/2]$. The two DWs become decoupled at $\tilde \alpha \rightarrow 0$, which occurs for either the grazing trajectory $\theta=\pm\pi/2$, or for $\tilde L\gg 1$. 
In this case, we recover the same shape for the ABS energy as that associated with the single DW configuration, see Eq.~(\ref{eq:E-bound}), but the dependence on $s(\theta)$ has been lost. 
The bound state energies for the DW located at $x=0$ correspond to $E_{0}(\theta)$, while those for the anti-wall at $x=L$ correspond to $-E_{0}(\theta)$, and consequently, our energy curves for the two DWs have two branches, as shown in the top panel of Fig.~\ref{fig:BSE2DW} in the case of $\chi = \pi$.  

We solve Eq.~(\ref{eq:BSE2DW}) numerically to obtain a profile for the bound state energies in the general case. To illustrate the effect of $\tilde L$ on the spectrum, we present the result for $\chi=\pi$ in Fig.~\ref{fig:BSE2DW}. As the distance between the DWs decreases, the bound states localized near them become hybridized, which leads to the splitting of the energy branches. 

\section{\label{sec:4} Interaction Between Domain Walls}

We are now in a position to evaluate the interaction energy between the DWs. We begin with the expression for the free energy of an arbitrary nonuniform superconducting texture in terms of the Fredholm determinant of the BdG Hamiltonian. 
For a system with two-component order parameter in zero magnetic field, it has the form:~\cite{Popov}
\begin{eqnarray}
	\label{eq:FreeE}
 	{\cal F} &=&  -T\sum_{n} \ln \mathrm{Det} \left( \frac{i \omega_{n}-H_{BdG}}{i \omega_{n}-H_{N}}\right)\nonumber\\
	 && + \frac{1}{V}\int \left( |\eta_{1}|^2 + |\eta_2|^2\right ) d^{2}r,
\end{eqnarray}
where ${\cal F}$ is the total free energy of the system measured with respect to the normal state with $\Delta(x)=0$, $H_{BdG}$ is the BdG Hamiltonian defined by Eqs.~(\ref{eq:BdGH}) and~(\ref{eq:BdG}), $H_{N}$ is the normal-state BdG Hamiltonian, $\omega_{n}=(2n+1)\pi T$ is the fermionic Matsubara frequency, and $V$ is the coupling constant in the chiral $p$-wave channel. 

We introduce the free energy difference, $\delta {\cal F}$, between nonuniform and uniform superconducting states, where the nonuniform state has two DWs. From Eq.~(\ref{eq:FreeE}), it follows that
\begin{eqnarray}
	\label{eq:dFreeE}
 	\delta	{\cal F} & = & -T\sum_{n} \ln \mathrm{Det} \left[ \frac{i \omega_{n}-H_{BdG}}{i \omega_{n}-H_{BdG}^{(0)}}\right]\nonumber\\ 
	  && + \frac{1}{V}\int \left( |\bm{\eta}|^2 - |\bm{\eta}^{(0)}|^2\right ) d^{2}r,
\end{eqnarray}
where $\bm{\eta}^{(0)}$ and $H_{BdG}^{(0)}$ denote the order parameter and the BdG Hamiltonian corresponding to the uniform chiral state. The expression~(\ref{eq:dFreeE}) depends on the separation between the DWs. Since the DWs are decoupled at infinite separation, $\delta {\cal F}(L \rightarrow \infty)$ gives the self-energy of two DWs, measured with respect to the uniform superconducting state. 
The interaction energy between the two DWs ${\cal F}_{int}$ is simply the difference between the free energy at arbitrary DW separation $L$ and the self-energy of the two-DW configuration, i.e. ${\cal F}_{int}=\delta {\cal F}(L)-\delta {\cal F}(L\rightarrow \infty)$.

For the sharp two-DW model introduced in Sec.~\ref{sec:2}, we have $|\bm{\eta}|^2=2\Delta_{0}^2$ in each of the three domains. Also, $|\bm{\eta}^{(0)}|^2=2\Delta_{0}^2$, and consequently the second term on 
the right-hand side of Eq.~(\ref{eq:dFreeE}) vanishes, and the interaction energy takes the form
\begin{equation}
	\label{eq:Fint}
 	{\cal F}_{int} = \tilde{\cal F}(L) - \tilde{\cal F}(\infty) 
\end{equation}
where 
\begin{equation}
	\label{eq:Ftilde}
	\tilde{\cal F}(L)=-T\sum_{n} \ln \mathrm{Det} \left[ \frac{i \omega_{n}-H_{BdG}(L)}{i \omega_{n}-H_{BdG}^{(0)}}\right] .
\end{equation}
The logarithm of each of the Fredholm determinants in Eq.~(\ref{eq:Fint}) can be written as follows:
\begin{equation}
	\label{eq:Det}
	\ln \mathrm{Det} \left[ \frac{i \omega_{n}-H_{BdG}}{i \omega_{n}-H_{BdG}^{(0)}}\right]=\sum_{i} \ln \left[ \frac{i \omega_{n}-E_{i}}{i \omega_{n}-E_{i}^{(0)}}\right],
\end{equation}
where $i$ is a set of quantum numbers labelling the eigenstates of the $2\times2$ BdG Hamiltonian, see Eq.~(\ref{eq:BdG}), at given DW separation $L$,  $E_{i}$ are the corresponding eigenvalues, and $E_{i}^{(0)}$ are the eigenvalues for the uniform chiral state. 

The sum over the BdG spectrum in Eq.~(\ref{eq:Det}) can be expressed in the semiclassical approximation as a sum over the eigenvalues of the Andreev Hamiltonian $H_{A}$, defined by Eq.~(\ref{eq:And}), as follows:
\begin{equation}
	\label{eq:Sum}
	\sum_{i}(\cdots)=2 \pi N_{F} \ell_{y} \int \frac{d \hat{\bm{k}}_F}{2\pi}|v_{F,x}|\sum_{j}(\cdots),
\end{equation}
where $N_{F}=m/2\pi$ is the density of states at the Fermi level per one spin projection in two dimensions, $\ell_{y}$ is the length of the DW, $\hat{\bm{k}}_F$ defines the direction of semiclassical propagation of the quasiparticles, 
and $j$ labels the eigenstates of the Andreev Hamiltonian at given $\hat{\bm{k}}_F$. Recall from section Sec.~\ref{sec:3a} that the quasiparticle wavefunctions satisfy $\psi(+\infty)=\psi(-\infty)$, so we have appropriately placed our system in a box of
dimensions $\ell_{x}=\ell$ and $\ell_{y}$ and imposed the periodic boundary conditions. 

It follows from Eqs.~(\ref{eq:Fint}, \ref{eq:Ftilde}, \ref{eq:Det}, \ref{eq:Sum}) that the interaction energy per unit DW length is given by:
\begin{equation}
	\label{eq:Fint2}
	F_{int}(L)=-2 \pi N_{F} T \sum_{n} \int_{0}^{2\pi} \frac{d\theta}{2 \pi} |v_{F,x}| \ln \frac{D(i \omega_{n}; L)}{D(i \omega_{n}; \infty)},
\end{equation}
with $D(z)=\prod_{j} (z - E_j)/(z - E_j^{(0)})$, where $E_j$ are the eigenvalues of the Andreev Hamiltonian at given $\hat{\bm{k}}_F$ for given DW separation $L$, and $E_{j}^{(0)}$ are the eigenvalues of the Andreev Hamiltonian in the uniform chiral state. Note that the gauge transformation introduced in Sec.~\ref{sec:3a} leaves the eigenvalues of $H_A$ unaffected. After the transformation, the Andreev Hamiltonian can be written as $H_{A}=H_{A}^{(0)}+\delta H$, where $H_{A}^{(0)} = -iv_{F,x}\hat{\sigma}_{3}\nabla_{x} + \Delta_{0}\hat{\sigma}_{1}$ is the Andreev Hamiltonian for the uniform chiral state which now has $\Delta_{\bm{k}_F}(x)=\Delta_{0}$, and $\delta H = v_{F, x}\Phi'(x)\hat{\sigma}_{0}/2$ is a localized perturbation, cf. Eq.~(\ref{eq:And-gt}). Adiabatically switching on the perturbation, there is a one-to-one correspondence between the eigenvalues of $H_{A}$ and $H_{A}^{(0)}$. 

Next we introduce the $2\times 2$ transfer matrix $\hat{M}(x; E)$ which acts as an $x$-evolution operator for the quasiparticle wavefunctions at given energy $E$: $\psi(x)=\hat{M}(x; E)\psi(-\ell/2)$. The transfer matrix satisfies the following conditions:
\begin{equation}
	\label{eq:Transfer}
	\left(H_{A}-E\right)\hat{M}(x; E) = 0, \quad  \hat{M}\left(-\ell/2;E\right)=\hat{\sigma}_{0},	 
\end{equation}
which hold for arbitrary values of $E$. It follows from the periodic boundary conditions that the quasiparticle wavefunctions satisfy $[\hat{\sigma}_{0}-\hat{M}(\ell/2; E)]\psi(-\ell/2)=0$. 
This quantization condition leads to the characteristic equation for the eigenvalues of $H_{A}$, given by $\det \left [ \hat{\sigma}_{0} - \hat{M} \left (\ell/2; E \right ) \right  ] = 0$, where $\det (\cdots)$ is a $2 \times 2$ determinant.
We also define another transfer matrix $\hat{M}^{(0)}(x;E)$, which satisfies the same conditions (\ref{eq:Transfer}), but for the uniform-state Hamiltonian $H_{A}^{(0)}$. 

From here we can introduce a new quantity $d(z)$, which is defined by the following expression:
\begin{equation}
	\label{eq:Fred2}
	d(z)=\frac{\det \left[ \hat{\sigma}_{0} - \hat{M}(\ell/2; z) \right ]} {\det \left[ \hat{\sigma}_{0} - \hat{M}_{0}(\ell/2; z) \right ]}.
\end{equation}
Both $d(z)$ and the Fredholm determinant $D(z)$ have zeros at $z=E_{j}$, as well as poles at $z=E_{j}^{(0)}$. For $|z|\rightarrow \infty$, which physically corresponds to large values of $E$, the quasiparticles are not affected by the superconducting order 
parameter. Consequently, $H_{A} \rightarrow H_{A}^{(0)}$, and $D(z),d(z)\rightarrow 1$. Due to these properties, we obtain: 
\begin{equation}
	\label{eq:Dd}
	D(z)=d(z),
\end{equation}
see, e.g. Ref.~\onlinecite{Dunne} for review.

In the following subsection we use the transfer matrix method, in particular Eq.~(\ref{eq:Dd}), to relate the DW interaction energy to the scattering matrix entries. Subsequently, we evaluate the sum over the Matsubara frequencies and the integral over semiclassical directions of propagation in Eq.~(\ref{eq:Fint2}) to obtain an analytical expression for the interaction energy in the limit of large DW separation.

\subsection{Calculation of the Fredholm determinant}

To facilitate the calculation of the Fredholm determinant at imaginary (Matsubara) energies, we first define a matrix $\hat \tau$, which relates the amplitudes of the gauge-transformed quasiparticle wavefunctions on the left-hand side of the DW configuration to those on the right-hand side, see Sec.~\ref{sec:3a}, as follows:
\begin{equation}
	\label{eq:tau}
	\left(\begin{matrix}
		A_{+}^{(+)}  \\ \noalign{\vskip 2mm} 
                     A_{-}^{(+)}
	\end{matrix}\right) = \hat \tau
	\left( \begin{matrix}
		A_{+}^{(-)}  \\ \noalign{\vskip 2mm} 
		A_{-}^{(-)}
	\end{matrix}\right).
\end{equation}
It can, therefore, be expressed in the following way:
\begin{equation}
	\label{eq:tau2}
	\hat \tau = \frac{1}{S_{22}}\left ( \begin{matrix}
		\det \hat S & S_{12}\\ \noalign{\vskip 2mm}
		-S_{21} & 1 \end{matrix}\right),
\end{equation}
where $\hat S$ is the scattering matrix defined by Eq. (\ref{eq:Scat}).
 
We introduce a shorthand notation for the transfer matrix from $-\ell/2$ to $+\ell/2$: $\hat{M}(\ell/2; z)=\hat{m}$. Using Eq.~(\ref{eq:tau}) and the wavefunctions defined by Eq.~(\ref{eq:QPwf}), we find that $\hat{m}= \hat{V}_{+} \hat{\tau} \hat{V}_{-}^{-1}$, where 
$$
\hat{V}_{\pm} = \left ( \begin{matrix}
	u_{+}e^{\pm i q \ell/2}  & u_{-} e^{\mp i q \ell /2} \\ \noalign{\vskip 2mm}
	e^{\pm i q \ell/2} & e^{\mp i q \ell /2} \end{matrix}\right),
$$
and $u_{\pm}$ and $q$ are defined in Sec.~\ref{sec:3a}.  We introduce a similar notation for the uniform-state transfer matrix $\hat{M}^{(0)}(\ell/2; z)=\hat{m}_{0}$, and since in the absence of DWs $\hat \tau = \hat S = \hat{\sigma}_{0}$, it immediately follows that $\hat{m}_{0}=\hat{V}_{+}\hat{V}_{-}^{-1}$. 

We can now rewrite the expression for the Fredholm determinant, see Eqs.~(\ref{eq:Fred2}) and~(\ref{eq:Dd}), as
\begin{equation}
	\label{eq:Fred3}
	D(z)=\frac{\det \left ( \hat{\sigma}_{0} - \hat{V}_{+}\hat \tau \hat{V}_{-}^{-1} \right )}{\det \left ( \hat{\sigma}_{0} - \hat{V}_{+}\hat{V}_{-}^{-1} \right )},
\end{equation}
and after multiplying the matrices we find that the numerator in the last equation takes the form 
\begin{equation}
	\label{eq:det}
	\det \left ( \hat{\sigma}_{0} - \hat{m} \right ) = 1 + \det \hat \tau - \left ( \tau_{11}e^{i q \ell} + \tau_{22} e^{-i q \ell} \right ).
\end{equation}
In the uniform superconducting state, this reduces to $\det (\hat{\sigma}_{0} - \hat{m}_{0}) = 2 ( 1 - \cos{q \ell} )$.

To calculate the interaction energy, we must evaluate $D(z)$ at discrete imaginary points $z=i\omega_{n}$, see Eq. (\ref{eq:Fint2}), and since $q$ in Eq.~(\ref{eq:det}) is only defined for real values of $E$ such that $|E| > \Delta_{0}$, we must analytically 
continue $q$ in the complex energy plane to the imaginary energy axis. Using the same procedure as in Sec.~\ref{sec:ABS} we find that the appropriate expression for $q$ is $q(E=i \omega_{n})=i\varkappa$, with 
$\varkappa=\sqrt{\omega_{n}^{2} + \Delta_{0}^{2}}/|v_{F,x}|$. Therefore the exponential terms in Eq.~(\ref{eq:det}) take the form $e^{\pm i q \ell}= e^{\mp \varkappa \ell}$, and in the thermodynamic limit $\ell \rightarrow \infty$, 
we keep only the last term in this equation as the others are small in comparison. In this limit, the Fredholm determinant in Eq.~(\ref{eq:Fred3}) becomes
\begin{equation}
	\label{eq:fred3}
	D(z=i \omega_{n})=\tau_{22}(i \omega_{n}) = \frac{1}{S_{22}(i \omega_{n})},
\end{equation}   
where we have used the relation between $\hat \tau$ and $\hat S$ given in Eq.~(\ref{eq:tau2}). Thus the calculation of the Fredholm determinant has been reduced to finding the properties of the scattering states. 

The expression (\ref{eq:fred3}) is applicable to any planar superconducting texture. In the case of two sharp DWs, using Eq. (\ref{eq:ScatEnt}), we obtain:
\begin{equation}
	\label{eq:S22}
	\frac{1}{S_{22}(i \omega_{n})}= 1 - \frac{\Delta_{0}^{2}\sin^{2}\delta}{\omega_{n}^{2}+\Delta_{0}^{2}}\left( 1 - e^{-2 \varkappa L}\right),
\end{equation}
where $\delta$ was defined in Sec. \ref{sec:3a}.

\subsection{\label{sec:4b} Interaction energy}

From Eqs.~(\ref{eq:Fint2}),~(\ref{eq:fred3}), and~(\ref{eq:S22}), it follows that the interaction energy per unit DW length at given DW separation has the following form:
\begin{eqnarray}
	\label{eq:Fint3}
	F_{int} & = & -v_{F} N_{F} T \sum_{n} \int_{0}^{2\pi} d\theta |\cos \theta|\nonumber\\
	    && \times \ln \left [ 1 + \frac{\Delta_{0}^{2} \sin^{2}\delta}{\omega_{n}^{2} + \Delta_{0}^{2} \cos^{2}\delta} e^{-2\varkappa L} \right ].
\end{eqnarray} 
The overall sign of the interaction energy is negative, so we see that the DW interaction mediated by the Andreev scattering of quasiparticles is attractive at all temperatures, which means there is an effective collapse of the walls 
to a single uniform domain. Qualitatively it is evident from Eq. (\ref{eq:Fint3}) that the attraction is exponentially weak in the limit of large separation. To make analytical progress, we focus on the case of zero temperature and $\chi=0$. 
The results for other values of the phase difference are expected to be qualitatively similar. 

At zero temperature, the summation over the discrete Matsubara frequencies $\omega_{n}$ becomes an integral over a continuous variable $\omega$: $T \sum_{n}(\cdots) \rightarrow \int(\cdots) d\omega/2\pi$. In the limit of large DW separation, 
one can expand the logarithm in Eq.~(\ref{eq:Fint3}), then the interaction energy takes the form:
\begin{eqnarray}
	\label{eq:Fint4}
	&& F_{int} = - \frac{N_{F} \Delta_{0} v_{F}}{2\pi} \int_{0}^{2\pi} d\theta |\cos \theta|\nonumber\\
	&& \times \int_{-\infty}^{+\infty} d\tilde \omega \frac{\sin^{2}\theta}{\tilde{\omega}^2 + \cos^{2}\theta} \exp\left(-\frac{2\tilde{L} \sqrt{\tilde{\omega}^{2}+1}}{|\cos \theta|}\right),
\end{eqnarray}
where $\tilde \omega = \omega/ \Delta_{0}$, and the dimensionless distance $\tilde L$ was introduced in Sec.~\ref{sec:ABS}. In the limit $\tilde L \gg 1$, one can further neglect the $\tilde\omega$-dependence of the pre-exponential factor 
in the integral over $\tilde \omega$, and evaluate this integral by the steepest descent method. In this way, we can represent Eq.~(\ref{eq:Fint4}) in the following form:
\begin{equation}
	\label{eq:Fint5}
	F_{int}= - \frac{2N_{F} \Delta_{0} v_{F}}{\sqrt{\pi \tilde{L}}} I(\tilde L) , 
\end{equation} 
where 
\begin{equation}
	\label{eq:tildeI}
	I(\tilde L) = \int_{0}^{\pi/2} d\theta \frac{\sin^{2} \theta}{\sqrt{|\cos \theta|}}  e^{-2\tilde{L}/|\cos \theta|},
\end{equation}
and we have invoked the symmetry of the angular integral to reduce the region of integration.

We make a change of variable $\rho=1/\cos\theta$ in Eq.~(\ref{eq:tildeI}) and obtain: $I = \int_{1}^{\infty} d\rho (\rho^{2}-1)^{1/2}  \rho^{-5/2}  e^{-2 \tilde L\rho}$.
This last integral can be expressed in terms of the modified Bessel functions of the second kind, and in the limit $\tilde L \gg 1$, it has the form $I=(10/3)\sqrt{\pi \tilde{L}}e^{-2\tilde{L}}$. Using this result in Eq.~(\ref{eq:Fint5}), and 
restoring the dimensional quantities, we arrive at our final expression for the interaction energy per unit DW length:
\begin{equation}
	F_{int}=-\frac{20}{3} N_{F}\Delta_{0} v_{F} \exp \left( -\frac{2\Delta_{0}L}{v_{F}} \right ).
\end{equation}
Thus, the interaction between the DWs is attractive and, as expected, it is exponentially weak in the limit of large separation between the walls.

\section{Conclusions}

We studied the interaction between two DWs separating states of opposite chirality in a $p$-wave superconductor, and found that it is attractive for arbitrary DW separation, at all temperatures. Furthermore, we found that the interaction energy is 
exponentially weak for large separation between the DWs. We used the transfer matrix method to relate the interaction energy of the DWs to the scattering matrix of the Bogoliubov quasiparticles, and the latter was calculated in the semiclassical (Andreev) approximation. 

The transfer matrix approach developed in this paper has a more general validity and can be applied to any superconducting texture. The free energy can be expressed in the same form as Eq.~(\ref{eq:dFreeE}) in terms of the Fredholm determinant of the BdG, 
or Andreev, Hamiltonian; however, the quasiparticle scattering matrix will be sensitive to the order parameter configuration. It would be interesting to use this method to characterize the interaction between DWs in a variety of other 
unconventional superconductors with discrete degeneracies of the ground states. 

Our results are immediately applicable to the neutral case. In a charged superconductor, there will be another contribution to the DW interaction coming from the Meissner currents and associated magnetic fields. Investigation of these effects on the interaction energy is left for the future work.

\begin{acknowledgments}
This work was supported by a Discovery Grant from the Natural Sciences and Engineering Research Council.
\end{acknowledgments}

\appendix

\section{\label{GL} GL Description of a Domain Wall}

To gain insight into the structure of the DW, we use the GL free energy functional. In a nonuniform neutral superfluid it is a sum of uniform ($F_{u}$) and gradient ($F_{g}$) energy densities. For the superconducting state with two-component order parameter $\bm{\eta}=(\eta_1,\eta_2)$, and isotropic Fermi surface, we have
\begin{equation}
	\label{GL-energy-u}
  	F_u=\alpha|\bm{\eta}|^{2}+\beta_1|\bm{\eta}|^{4}+\beta_2|\bm{\eta}\cdot \bm{\eta}|^2, 
\end{equation}
and
\begin{eqnarray}
	\label{GL-energy-g}
  	F_g & = & K_{1}(\nabla_i \eta_j)^*(\nabla_i \eta_j)+K_2(\nabla_i \eta_i)^*(\nabla_j \eta_j)\nonumber\\
	&& + K_3(\nabla_i \eta_j)^*(\nabla_j \eta_i),
\end{eqnarray}
where the Einstein summation convention is assumed. We find that the minimum in the uniform free energy density corresponds to the chiral states $\bm{\eta}=\Delta_{0}(1,\pm i)$ with $\Delta_{0}=\sqrt{|\alpha|/4\beta_{1}}$, for the case $\beta_1$, $\beta_2 > 0$.   

The order parameter of a planar DW configuration can be expressed in the following form:
\begin{equation}
	\label{eq:OPf}
	\begin{aligned}
	&\eta_{1}(x)=\Delta_{0}f_{1}(x)e^{i\phi(x)},\\
	&\eta_{2}(x)=\Delta_{0}f_{2}(x)e^{i\phi(x)-i\gamma(x)},
	\end{aligned}
\end{equation}
where $f_{1,2}$ are the dimensionless amplitudes of the order parameter components, whose asymptotics far from the DWs are given by $f_{1,2}=1$.

The nonzero phase difference $\chi$ between domains of opposing chirality emerges from the condition of vanishing supercurrent across the DW. It follows from Eq.~(\ref{GL-energy-g}) that the supercurrent is given by the expression $j_{i}=2\im\,(K_{1}\eta_{j}^{*}\nabla_{i}\eta_{j}+K_{2}\eta_{i}^*\nabla_{j}\eta_{j}+K_{3}\eta_{j}^{*}\nabla_{j}\eta_{i})$. Using Eq.~(\ref{eq:OPf}), we find that the transverse current has the form $j_{x}=2\Delta_0^2(K_{123}f_1^2+K_1f_2^2)(\nabla_x\phi)-2K_1\Delta_0^2f_2^2(\nabla_x\gamma)$, with $K_{123}=K_1+K_2+K_3$.

In our work we focus on the static case, so conservation of current requires $j_{x}=\mathrm{const}$. One can set $j_{x}=0$, which yields a linear relation between the gradients of $\phi$ and $\gamma$. This relation can be used to eliminate the common phase 
from the gradient energy and obtain:
\begin{equation}
    \label{F-ug-reduced}
    \begin{array}{rcl}
  	F_u&=&\alpha\Delta_0^2(f_1^2+f_2^2)+\beta_1\Delta_0^4(f_1^2+f_2^2)^2 \medskip \\ 
	&&+\beta_2\Delta_0^4(f_1^4+f_2^4+2f_1^2f_2^2\cos 2\gamma),\\ \\
  	F_g&=&K_{123}\Delta_0^2(\nabla_xf_1)^2+K_1\Delta_0^2(\nabla_xf_2)^2  \medskip \\
	&&+\dfrac{K_1K_{123}f_1^2f_2^2}{K_{123}f_1^2+K_1f_2^2}\Delta_0^2(\nabla_x\gamma)^2.
   \end{array}
\end{equation}
Variational minimization of these expressions with respect to $f_{1,2}$ and $\gamma$ gives rise to three coupled nonlinear differential equations. Using the solutions to these equations, one can compute the common phase difference between
arbitrary points $x_1$ and $x_2$:     
\begin{equation}
	\label{phi-gamma}
	\phi(x_2)-\phi(x_1)=\int_{x_1}^{x_2}\frac{K_1f_2^2}{K_{123}f_1^2+K_1f_2^2}(\nabla_x\gamma)\,dx.
\end{equation}
We see that whenever there is a gradient of the relative phase $\gamma$, the value of the common phase difference is nonzero, and it is evidently sensitive to the microscopic details of the system. 
While we have imposed the condition of zero transverse current to obtain Eq.~(\ref{phi-gamma}), the current along the DW is nonzero.

Obtaining an exact analytical solution for the DW structure is not possible due to the complexity of the differential equations obtained from variational minimization of the expressions in Eq.~(\ref{F-ug-reduced}). However, one can make analytical progress 
by considering the constant-amplitude model,~\cite{VG85} in which case $f_{1,2}(x)=1$ for all $x$. Then it follows from Eq.~(\ref{F-ug-reduced}) that the total free energy density is given by the expression 
$F=F_u+F_g = (\cdots) +\tilde K\Delta_0^2(\nabla_x\gamma)^2+2\beta_2\Delta_0^4\cos 2\gamma$, where $\tilde K=K_1K_{123}/(K_{123}+K_1)$. The first term in $F$ contains $\gamma$-independent contributions, and it is immediately clear 
that the equation for $\gamma$ has a sine-Gordon form:
\begin{equation}
	\label{eq:SG}
	\tilde K\nabla_x^2\gamma+2\beta_2 \Delta_{0}^2\sin{2\gamma}=0.
\end{equation}
The simplest nontrivial solution to this equation is a kink-like solution given by $\sin\gamma(x)=\tanh(x/\xi_d)$. This corresponds to a single DW. 
The parameter $\xi_d=\sqrt{\tilde K/4\beta_2\Delta_0^2}$ has the physical meaning of the DW thickness and is of the order of the GL correlation length. Using this expression for $\gamma(x)$ in Eq. (\ref{phi-gamma}) we obtain:
$\chi\equiv\phi(+\infty)-\phi(-\infty)=\pi K_1/(K_{123}+K_1)$. In the weak coupling model, $K_1=K_2=K_3$ (Ref. \onlinecite{Book}), and $\chi=\pi/4$.

There are no stable two-kink (or two-DW) solutions to Eq.~(\ref{eq:SG}); the other nontrivial solution corresponds to a periodic lattice of DWs. This can be understood by considering a simple pendulum with $2\gamma \rightarrow \Theta$ 
(the angular displacement of the pendulum), and $x \rightarrow t$ (a time coordinate). Consider first the one-kink solution, where initially $t \rightarrow -\infty$, and we have $\Theta = -\pi$. After a sufficient amount of time has elapsed, 
the pendulum has just enough energy to complete one full revolution, approaching an angular displacement of $\Theta=+\pi$. There can be no two-kink solutions because if the pendulum has enough energy to surpass the limit $\Theta=+\pi$ and 
complete one more revolution, it will have enough energy to do this an infinite number of times, which corresponds to a periodic arrangement of DWs.    

The fact that there are no stable two-DW solutions implies there must be some form of interaction between the walls. If this interaction is attractive, it will cause a collapse 
of the DWs to a single domain. If it is repulsive, then one of the DWs will be pushed to infinity, leaving just a single DW separating two domains of opposite chirality. 
The periodic solution for $\gamma(x)$ can also be understood in terms of this interaction, as a mutual attraction or repulsion between neighbouring DWs could potentially lead to a stable periodic configuration.

\end{document}